\documentclass{pasj00}
\draft

\begin{document}
\SetRunningHead{Takizawa et al.}{Mass Estimation of Merging Galaxy Clusters}
\Received{yyyy/mm/dd}%{yyyy/mm/dd}
\Accepted{yyyy/mm/dd}%{yyyy/mm/dd}

\title{Mass Estimation of Merging Galaxy Clusters}

%%% begin:list of authors
% Do NOT capitalize all letters in "textsc".
%\author{A-Firstname \textsc{A-Familyname} %
%  \thanks{Example: Present Address is xxxxxxxxxx}}
%\affil{A-Address of Institute}
%\email{aaaaa@xxx.xxx.xx.xx}

%\author{B-Firstname \textsc{B-Familyname}}
%\affil{B-Address of Institute}\email{bbbbb@xxx.xxx.xx.xx}
%\and
%\author{C-Firstname {\sc C-Familyname}}
%\affil{C-Address of Institute}\email{ccccc@xxx.xxx.xx.xx}
%%% end:list of authors

%%% Please use the following style in case that sorting by 
%%% affilation is impossible. 
%
 \author{%
   Motokazu \textsc{Takizawa}\altaffilmark{1}
   Ryo \textsc{Nagino}\altaffilmark{2}
   and
   Kyoko \textsc{Matsushita}\altaffilmark{2}}
 \altaffiltext{1}{Department of Physics, Yamagata University, 1-4-12 Kojirakawa-machi,
       Yamagata 990-8560}
 \email{takizawa@sci.kj.yamagata-u.ac.jp}
 \altaffiltext{2}{Department of Physics, Tokyo University of Science, 1-3
      Kagurazaka, Shinjuku-ku, Tokyo 162-8601}

%% `\KeyWords{}' always has to be placed before `\maketitle'.
 %Do NOT move this preamble from here!

\maketitle

\begin{abstract}
We investigate the impact of mergers on the mass estimation of galaxy clusters
using $N$-body + hydrodynamical simulation data. We estimate virial mass
from these data and compare it with real mass. 
When the smaller subcluster's mass is larger than a quarter of that 
of the larger one, virial mass can be larger than twice of the real mass.
The results strongly depend on the observational directions, because of 
anisotropic velocity distribution of the member galaxies.
We also make the X-ray surface brightness and spectroscopic-like temperature
maps from the simulation data. The mass profile is estimated from these data 
on the assumption of hydrostatic equilibrium. In general, mass estimation 
with X-ray data gives us better results than virial mass estimation. 
The dependence upon observational directions is weaker than in case of virial 
mass estimation. When the system is observed along the collision axis, 
the projected mass tends to be underestimated. This fact should be noted 
especially when the virial and/or X-ray mass are compared with 
gravitational lensing results.
\end{abstract}

\section{Introduction}
Mass is one of the most important physical parameters to characterize
astrophysical objects. This is especially true in many kinds of
self-gravitating objects such as star clusters, galaxies, and clusters
of galaxies, where gravity plays a crucial role in their evolution.
In addition, mass distribution in large scales in the
universe have some information about dark matter (DM). For example, mass
distribution in the very central part of relaxed dark halos could be altered
by the self-interaction of DM particles \citep{Yosh00}. 
Investigating spatial distributions of DM, galaxies, and intracluster
medium (ICM) also gives us important clues of DM properties \citep{Clow04,
Brad06, Jee07, Okab08}.

There are several methods to estimate mass distribution of galaxy
clusters in an observational way. Assuming dynamical equilibrium,
we can estimate cluster mass from line-of-sight velocity distribution of
member galaxies through virial theorem or Jeans equation 
\citep{Kent82, Oege94, Gira98, Barr07, Maur08}.
Cluster mass distribution can also be estimated from the X-ray
observational data on the assumption of hydrostatic equilibrium of the
ICM \citep{Reip02, Taki03, Ota04, Vikh06, Gast07}.
Weak and strong gravitational lensing techniques enable us to determine
gravitational mass directly without any assumptions of dynamical
status of the system \citep{Kais93, Mira95}. 

We have multiple observational ways of cluster mass estimation as
mentioned above. However, these methods do not always give us
consistent results \citep{Mahd08}. In case of well-known gravitational
lensing cluster CL 0024+17, for example, mass derived from X-ray data
\citep{Ota04} and those from gravitational lensing \citep{Tyso98, Broa00, Jee07} 
are significantly different, where the latter results are larger than the former
by a factor of two or three.
Recently, \citet{Zhan10} made a comparison of X-ray and lensing mass
for 12 clusters, and showed that mass of disturbed clusters could be both overestimated
and underestimated.

Some assumptions are necessary in the mass estimation method presented
above. In the method with line-of-sight galaxy velocities, we assume
that the system is in dynamical equilibrium and often assume 
isotropic velocity distribution of the member galaxies and spherical
symmetric structure. In addition, another critical aspect in this method lies in the 
difficulty of removing foreground and background galaxies, which significantly
affect the estimate of the velocity dispersion and mass \citep{Bivi06}.
In the X-ray method we assume that the systems are in
hydrostatic equilibrium and spherically symmetric. Simple methods with
strong lensing often assume axial symmetry. Even more elaborate methods
based on strong lensing technique are essentially model-dependent.
Although the weak gravitational lensing method seems to be free from these
assumptions about dynamical state and geometry, it does not provides us with
three-dimensional mass density distribution but two-dimensional
surface-mass density one. This fact should be noted especially 
when the lensing results are compared with those derived from other methods.

A lot of $N$-body + hydrodynamical simulations with cosmological initial conditions
have been done so far to investigate accuracy of cluster mass estimation
\citep{Evra96, Rasi06, Kay07, Naga07, Piff08, Vent08, Ameg09}.
It is true that such studies are useful in order to investigate this issue in realistic 
situations. However, such simulations are not suitable to model a single merger event 
because it is difficult to control initial conditions and achieve very high resolution 
in particular interesting regions efficiently. 
Therefore, studies based on binary merger simulations are complementary and worthwhile,
in which it is relatively easy to keep initial conditions under control, improve numerical resolution,
and derive clear physical interpretations.
The assumptions for mass estimation listed above are not very
good in clusters during or a few Gyr after mergers. Binary merger cluster simulations 
tell us that mergers generate bulk flow motion and complex temperature structures
in the ICM, and anisotropic velocity distribution in DM and
galaxies \citep{Roet96, Taki99, Taki00, Rick01, Ritc02, Rowl04, Asca06, Pool06,
Taki06, McCa07, Pool07, Spri07, Akah08, Mast08, ZuHo09}. 
Of course, asymmetric spatial structures are also often seen in ICM, DM, and galaxies.
However, it is not trivial how these systems will be
overestimated or underestimated. This depends on the phase of mergers,
geometry of the system and observational direction, and which mass
estimation method we use.

In this paper, we adopt the latter approach to 
investigate the impact of mergers on the cluster mass estimation
using $N$-body+hydrodynamical simulation data. We make mock
observational data such as line-of-sight velocity of member galaxies, 
X-ray brightness distribution, and spectroscopic-like temperature \citep{Mazz04} maps
of ICM from the simulation data. We perform ``simulations of mass
estimation'' for these mock data, and compare the results with actual mass
distribution in the simulation data.

The rest of this paper is organized as follows. In \S \ref{s:simu} we
describe the adopted numerical methods and initial conditions for our
$N$-body + hydrodynamical simulations. In \S \ref{s:resu} we present the
results of mass estimation. 
In \S \ref{s:summ} we summarize the results and discuss their implications.

\section{The Simulations}
\label{s:simu}

\subsection{Numerical Methods}
Numerical methods used here are basically the same as those in
\citet{Taki06}, whose hydrodynamical part is identical with what was 
used in \citet{Taki05}. 
In the present study, we consider clusters of galaxies
consisting of two components: collisionless particles corresponding to
the galaxies and DM, and ideal fluid corresponding to the ICM. 
When calculating gravity, both
components are considered, although the former dominates over the
latter. Radiative cooling and heat conduction are not included.
We use the Roe total variation diminishing (TVD) scheme to solve the
hydrodynamical equations for the ICM (see \cite{Hirs90}). 
The Roe scheme is a well-known Godunov-type method with a linearized Riemann solver \citep{Roe81}. 
It is relatively simple and good at capturing shocks without any artificial viscosity. 
Using the MUSCLE approach and a minmod TVD limiter, we obtain second-order accuracy 
without any numerical oscillations around discontinuities. To avoid negative pressure, 
we solve the equations for the total energy and entropy conservation simultaneously. 
This method is often used in astrophysical hydrodynamic simulations 
where high Mach number flow can occur \citep{Ryu93, Wada01}.
Gravitational forces are calculated by the
Particle-Mesh (PM) method with the standard Fast Fourier Transform
technique for the isolated boundary conditions (see \cite{Hock88}).
The size of the simulation box is $(9.40 \,\mathrm{Mpc}) \times (4.70 \,\mathrm{Mpc})^2$.
The number of the grid points is $256 \times (128)^2$.
The total number of the $N$-body particles used in the simulations is also 
$256 \times (128)^2$, which is approximately $4.2 \times 10^6$. 

It is certain that simulations about full history of realistic cluster formation and 
evolution should include 
gas cooling, star formation and energy feedback. 
However, the aim of the paper is to investigate the impact of merger
dynamics on the cluster mass estimation in rather idealized conditions in order
to clarify physical interpretation. For this reason, we restrict ourselves to
consider simulations which do not incorporate radiative cooling, which
is dynamically significant only in the cluster core regions.

\subsection{Models and Initial Conditions}

The equilibrium cluster model used here is essentially the same as what 
is used in \citet{Taki08} except that there is no magnetic field. Full description
is written in section 2 of \citet{Taki08}. 
We consider mergers of two virialized subclusters with an NFW density
profile \citep{Nava97} in the $\Lambda$CDM universe
($\Omega_0=0.25$, $ \lambda_0=0.75$) for DM.
Given the cosmological parameters and halo's virial mass,
we calculate the parameters in the NFW profile following a method in Appendix
of Navarro, Frenk, White (1997).
The initial density profiles of the ICM are assumed to be
those of a beta-model. We assume that the core radius is half of the
scale radius of the DM distribution, and that $\beta=0.6$. The gas mass
fraction is set to be 0.1 inside the virial radius of each subcluster.
The velocity distribution of
the DM particles is assumed to be an isotropic Maxwellian. The radial
profiles of the DM velocity dispersion are calculated from the Jeans
equation with spherical symmetry, so that the DM particles would be in
virial equilibrium in the cluster potential of the DM and ICM.
The radial profiles of the ICM pressure are determined in a similar
way so that the ICM would be in hydrostatic equilibrium within
the cluster potential with a plausible boundary condition. 
Because the gas density profiles are already given, 
temperature profiles are automatically determined with an equation of state of
the ideal gas.
Strictly speaking, the above-mentioned model is in dynamically equilibrium
when the spatial distribution of DM and ICM extend to the infinity.
In addition, finite spatial resolution of the code 
could affect the structure and evolution especially in the central region.
These issues are discussed in Appendix.

Given each subcluster's mass and an angular momentum parameter of the system,
we estimate ``typical'' initial conditions for cluster mergers
in the same way of \citet{Taki08}.
%%%%%%%%%%%%%%%%
We restrict ourselves to analyze head-on mergers because their simple geometrical
structures are suitable for the purpose of this paper to clarify physical interpretation 
of the phenomena. We believe that this choice is fairly reasonable considering that the distribution of 
impact parameters might be biased to lower values if most mergers occur along large scale structure 
filament. 
Off-center merger cases are certainly interesting and important especially
to explore more general situations and/or investigate particular observational results,
which will be investigated as a future work.
%%%%%%%%%%%%%%%
How to estimate these conditions is described in detail in Appendix of \citet{Taki08},
where the initial relative velocity and impact parameter are calculated through energy
and angular momentum conservation laws and some well-known scaling relations predicted from
a spherical collapse model (see \cite{Peeb80}).
Simulations start in condition that the both subclusters touch each other.
Typically, relative initial velocity becomes about two thirds of the
infall velocity assuming that they were at rest at infinite distance.
%%%%%%%%%%%%%%%%%
This value is similar to the circular velocity of the larger halo at the virial radius.
Thus, our initial conditions are quite equivalent to those of controlled merger simulations in
\citet{Pool06} and \citet{McCa07} that are constructed so as to reproduce the merger events
found in cosmological simulations.
%%%%%%%%%%%%%%%%%%
The coordinate system is
taken in such a way that the center of mass is at rest at the origin.
Two subclusters are initialized in the $xy$-plane.
The collision axis is along the $x$-axis.
The centers of the larger and smaller subclusters were initially
located at the sides of $x<0$ and $x>0$, respectively.

We analyze the simulation data of head-on collisions whose mass ratios are
1:1, 2:1, 4:1, 8:1, and 16:1. In general, qualitative features are roughly common to 
all models. Thus, we choose 4:1 merger run as a representative example, 
whose results will be investigated and described the most intensively.
The parameters for each model are summarized in table 1, where $M_i$, 
$r_{{\rm vir},i}$, $c_i$, $N_i$, $v_{\rm rel.ini}$ are the total mass, virial radius, 
concentration parameter, number of $N$-body particles for the $i$-th
subcluster, and initial relative velocity, respectively.

 \begin{table}
  \caption{Model Parameters}
  \label{tab:modpar}
  \begin{center}
   \begin{tabular}{cccccc}
    \hline \hline 
        &  $M_1/M_2 (10^{14} \mbox{\ensuremath{{M}_{\odot}}})$  &  
           $r_{{\rm vir},1}/r_{{\rm vir},2}$ (Mpc)  &  $c_1/c_2$ & $N_1/N_2$ & $v_{\rm rel.ini}$ (km s$^{-1}$)\\
    \hline
  Run1:1      & 5.0/5.0    & 1.57/1.57  & 6.12/6.12   &  2097152/2097152   &  1360\\
  Run2:1      & 5.0/2.5    & 1.57/1.24  & 6.12/6.60   &  2796202/1398102   &  1249\\
  Run4:1      & 5.0/1.25   & 1.57/0.987 & 6.12/7.08   &  3355443/838861    &  1210\\
  Run8:1      & 5.0/0.625  & 1.57/0.784 & 6.12/7.56   &  3728270/466034    &  1213\\
  Run16:1     & 5.0/0.3125 & 1.57/0.622 & 6.12/8.04   &  3947580/246724    &  1239\\
    \hline
   \end{tabular}
  \end{center}
%Note. -- This is a note.
 \end{table}

\section{Results}
\label{s:resu}

\subsection{Mass Estimation through Virial Theorem}
We ``observe'' the cluster in the simulation data from a certain direction.
Although we can use more than a million particles' velocity data,
it is unrealistic at present to observe such a huge number of member galaxy's 
line-of-sight velocities for a single cluster.
Thus, we randomly choose $N_{\rm samp}$ particles from the $N$-body
ones within a circle on the sky plain whose center and radius are mass center on the sky plain and
a virial radius of the larger cluster listed in table 1, respectively, 
and recognize them as galaxies whose line-of-sight velocity is
observed. Virial mass is calculated following the equation,
\begin{eqnarray}
  M_{\rm VT} = \frac{3 \pi}{G} \sigma^2_{\rm los} 
               \biggl\langle \frac{1}{R} \biggr\rangle^{-1},
\end{eqnarray}
where $\sigma^2_{\rm los}$ and $G$ are the dispersion of line-of-sight
velocities and gravitational constant, respectively. 
$\langle 1/R \rangle^{-1}$ is the
harmonic mean of the distance projected on the sky plain 
for particle pairs and defined as,
\begin{eqnarray}
\biggl\langle \frac{1}{R} \biggr\rangle^{-1} \equiv N_{\rm pair}
   \sum_{i>j}^{N_{\rm pair}} \frac{1}{R_{ij}},
\end{eqnarray}
where $R_{ij}$ is the distance projected on the sky plain between
$i$ and $j$-th ``observed'' galaxies. 
$N_{\rm pair} = N_{\rm samp}(N_{\rm samp}-1)/2$ is the number of pairs of
``observed'' galaxies. 
Please note that both spherical symmetry and isotropic
velocity dispersion are explicitly assumed in this method (see \cite{Binn87}).
We estimate virial mass using different $N_{\rm tr}$ sets of
``member galaxies'' for a single combination of a particular simulation
data snapshot and observed direction, and then 
calculate mean and variance of the virial mass.

\begin{figure}
  \begin{center}
    \FigureFile(80mm,80mm){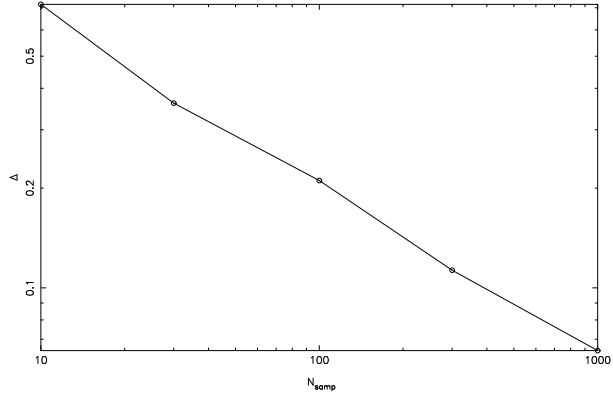}
    %%% \FigureFile(width,height){filename}
  \end{center}
  \caption{Dependence of statistical errors of virial mass upon the number of the
         member galaxy whose line-of-sight velocity can be used in the
         mass estimation.}\label{fig1}
\end{figure}

In actual observations to determine cluster mass, it is typical that 
$\sim 100$ member galaxy's line-of-sight velocities are available.
Even only a few tens of galaxies are sometimes used. Therefore, even if
the distributions of galaxies and DM are the same in the phase space, 
the spatial extent and velocity dispersion of the galaxies
sampled sparsely can be significantly different from those of DM.
To investigate this, we estimate virial mass for a single relaxed
cluster, which is identical to those in the initial conditions of merger
simulations, with $N_{\rm tr}=100$ and various $N_{\rm samp}$. 
Figure \ref{fig1} shows the
dependence of the virial mass 1$\sigma$ errors upon $N_{\rm samp}$. Clearly,
the errors are proportional to the $1/\sqrt{N_{\rm samp}}$. The relation is
approximately $\Delta \sim 0.2 N_{\rm samp}^{-1/2}$, which means that a
few ten percents errors are expected in usual cases 
($N_{\rm samp} \simeq 100$). 
It should be noted that another systematic errors could occur when
galaxies and DM obeys significantly different distribution.

\begin{figure}
  \begin{center}
    \FigureFile(80mm,80mm){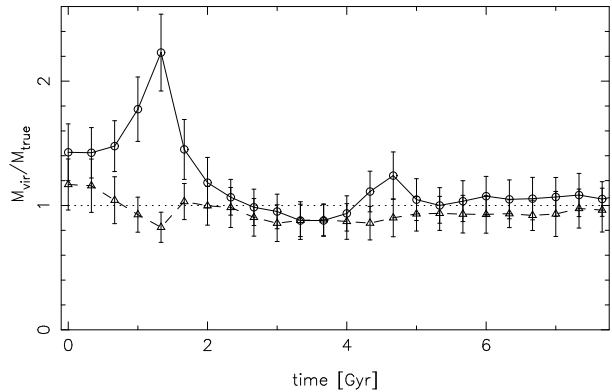}
    %%% \FigureFile(width,height){filename}
  \end{center}
  \caption{Evolution of the ratio of virial mass to ``real mass'' within
         the sphere for the 4:1 merger.
         Circles with solid lines and triangles with dashed lines
         represent results observed from the direction parallel and 
         perpendicular to the collision axis, respectively.
         Error bars represent
        1$\sigma$ errors in 100 trials of the estimation.}\label{fig2}
\end{figure}

\begin{figure}
  \begin{center}
    \FigureFile(160mm,160mm){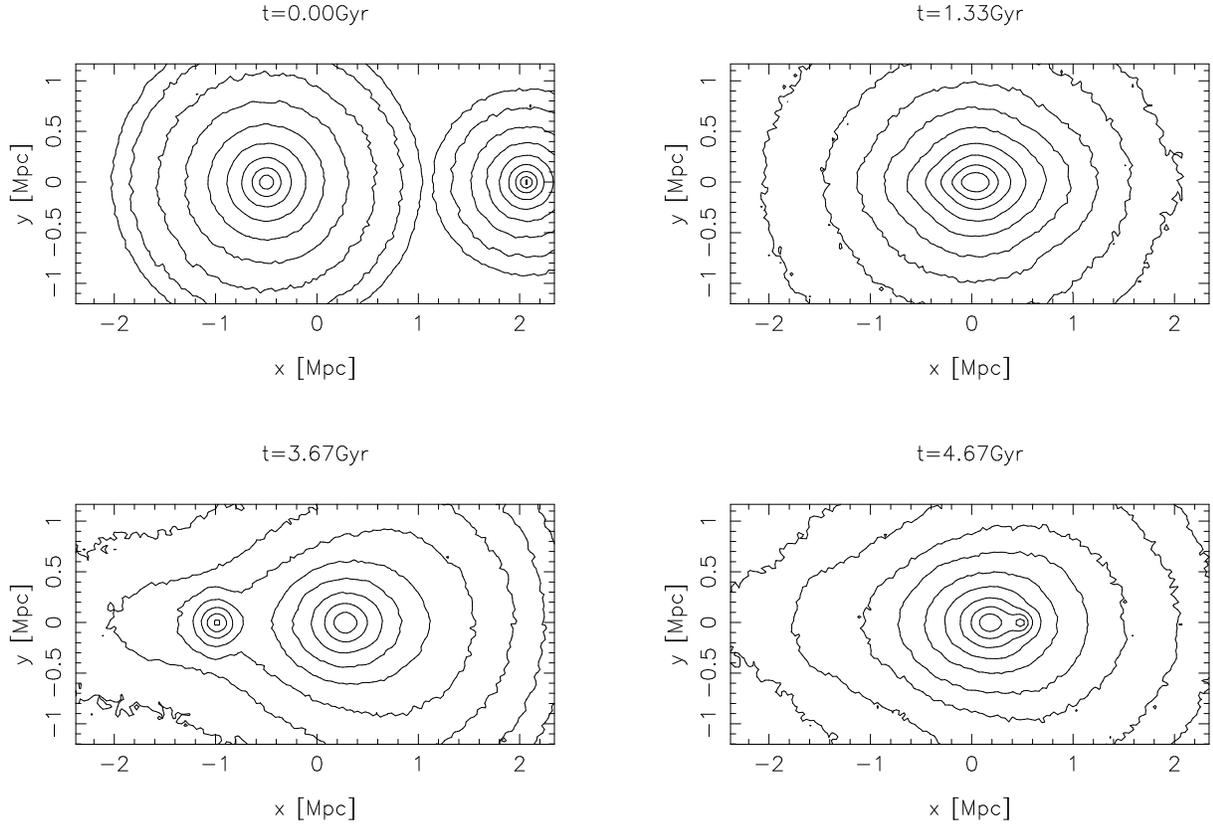}
    %%% \FigureFile(width,height){filename}
  \end{center}
  \caption{Snapshots of projected mass distribution for the 4:1 merger model
         seen from the direction perpendicular to the collision axis.}\label{fig3}
\end{figure}

We estimate virial mass in a similar way for merger simulation data
with $N_{\rm tr}=100$ and $N_{\rm samp}=100$, and compare the results to the real
mass. 
However, it is not so trivial what is ``real mass'' that should be
compared with because the system can be non-spherical. Considering that
spherical symmetry is assumed in the estimation, mass within a sphere
whose center corresponds to the mass center may be valid in theoretical
point of view. 
Figure \ref{fig2} presents evolution of the ratio of virial
mass to ``real mass'' within the sphere for 4:1 merger.
Circles with solid lines and triangles with
dashed lines represent results observed from the direction parallel and
perpendicular to the collision axis, respectively. Error bars represent
1$\sigma$ errors in $N_{\rm tr}$ trials of the estimation.
Snapshots of mass distribution for this model seen from the direction
perpendicular to the collision axis at representative epochs
are presented in figure \ref{fig3}. 

Here, we describe the comparison of virial mass with real mass within the
sphere. When the system is observed along the collision axis, mass tends
to be overestimated except during a short period around the apocenter of
the subcluster ($t\simeq3.7$Gyr). The most serious overestimation is
seen at the first core passage ($t\simeq1.3$Gyr). The second core
passage also cause the second but smaller maximum of overestimation
($t\simeq4.6$Gyr). Even after that, mass is systematically overestimated,
though this is within the statistical errors. When the system is observed in the
direction perpendicular to the collision axis, on the other hand, mass
tends to be underestimated except around the initial states. 
Again, the most serious underestimation occurs at the first core passage.
However, the extent of systematic underestimation after that is weak,
which is within the statistical errors. It is interesting that the
weak dependence of virial mass on the observed direction still remains
$\sim 6$ Gyr after the core passage.

\begin{figure}
  \begin{center}
    \FigureFile(80mm,80mm){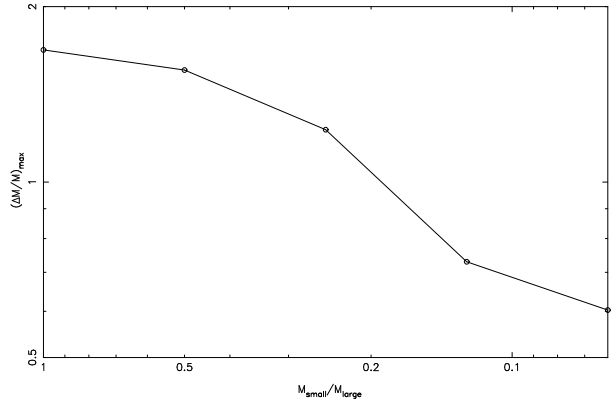}
    %%% \FigureFile(width,height){filename}
  \end{center}
  \caption{Dependence of maximum systematic errors of virial mass upon the
         mass ratio when observed along the collision axis. 
         }\label{fig4}
\end{figure}

As we wrote above, the largest systematic errors of virial mass
estimation are seen at the first core passage when the system is 
observed along the
collision axis. Naturally, it is expected that this depends on the mass
ratio of the smaller subcluster to the larger one. Figure \ref{fig4} shows
the dependence of the maximum systematic errors of virial mass upon the mass
ratio when observed along the collision axis. 
Roughly speaking, the mass of the system can be overestimated by a factor of more than two
when the smaller subcluster's mass is larger than a quarter of the larger one.

\subsection{Mass Estimation with X-ray Data}
Again, it is assumed that we ``observe'' the cluster in the simulation
data from a certain direction. We make two-dimensional 
X-ray brightness distribution and spectroscopic-like temperature \citep{Mazz04} maps
on the sky plain from the three-dimensional simulation data. 
We determine the image centers from the X-ray brightness
maps, and make one-dimensional profiles of the X-ray surface brightness
$I_{\rm X}(R)$ and spectroscopic-like temperature $T_{\rm sl}(R)$,
where $R$ is the distance projected on the sky plain from the image
center. The radial density profiles $\rho_g(r)$, where $r$ is the radial
distance, are calculated in a standard deprojection procedure assuming
spherical symmetry. This deprojection procedure is not done for
temperature and we use $T_{\rm sl}$ for mass estimation.
Both $\rho_g(r)$ and $T_{\rm sl}(r)$ are fitted by the $\beta$-model function.
If a single $\beta$-model function fit is not acceptable, we use a so-cold double
$\beta$-model, which is represented by a sum of two independent $\beta$ model functions.
Assuming hydrostatic equilibrium, the mass profile derived from the
X-ray observation is,
\begin{eqnarray}
  M_{\rm X}(r) = - \frac{k_{\rm B} T_{\rm sl} r}{G \mu m_{\rm p}}
                 \biggl(
    \frac{d \ln \rho_{\rm g}}{d \ln r} + \frac{d \ln T_{\rm sl}}{d \ln r}
                 \biggr),
    \label{eq:xrmass}
\end{eqnarray}
where $k_{\rm B}$, $\mu$, and $m_{\rm p}$ are the Boltzmann constant,
mean molecular weight, and proton mass, respectively.
It should be noted that the double $\beta$-model function shows
rather small gradient around an intersection of each $\beta$-model
function. This possibly causes unphysical structures in the estimated mass
profile through equation (\ref{eq:xrmass}).

\begin{figure}
  \begin{center}
    \FigureFile(160mm,160mm){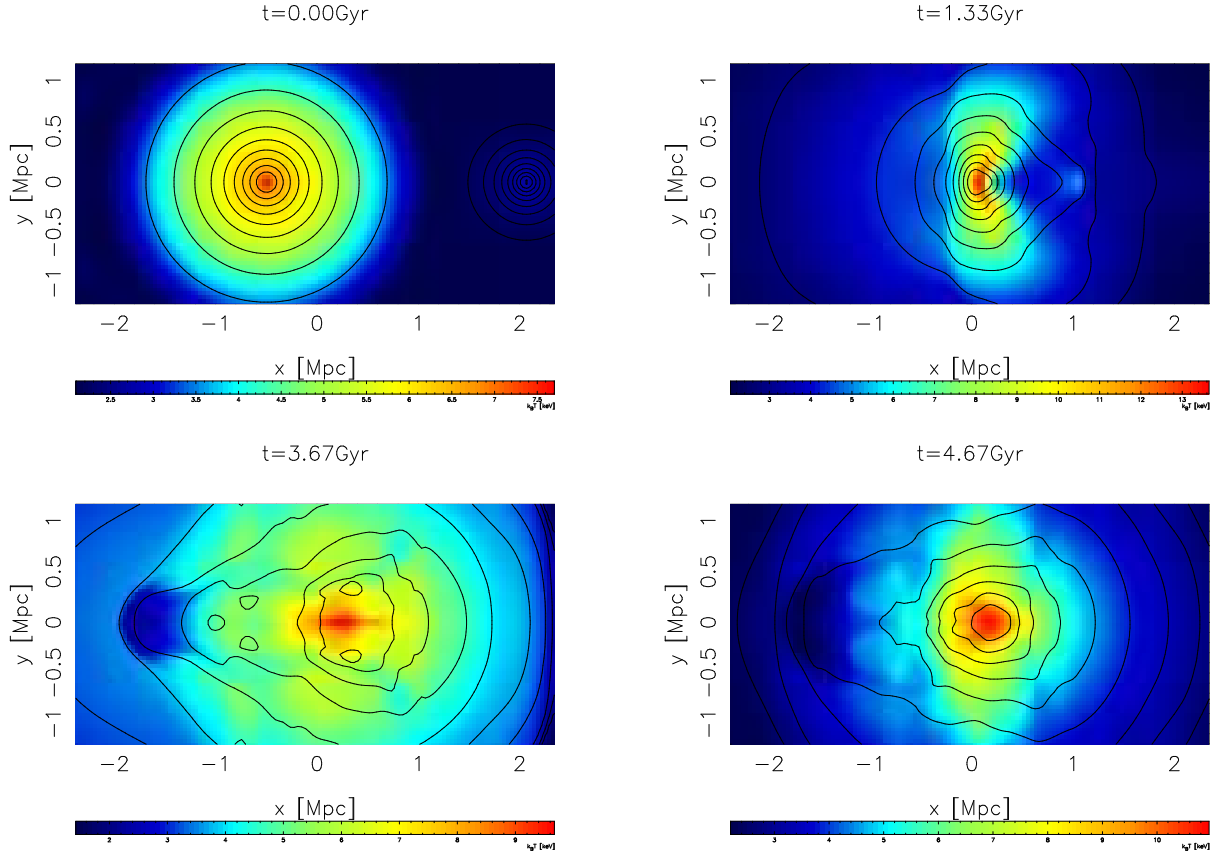}
    %%% \FigureFile(width,height){filename}
  \end{center}
  \caption{Snapshots of X-ray surface brightness distribution (contours)
         and spectroscopic-like temperature map (colors) for the 4:1
         merger model seen from the direction perpendicular to the
         collision axis.}\label{fig5}
\end{figure}

\begin{figure}
  \begin{center}
    \FigureFile(160mm,160mm){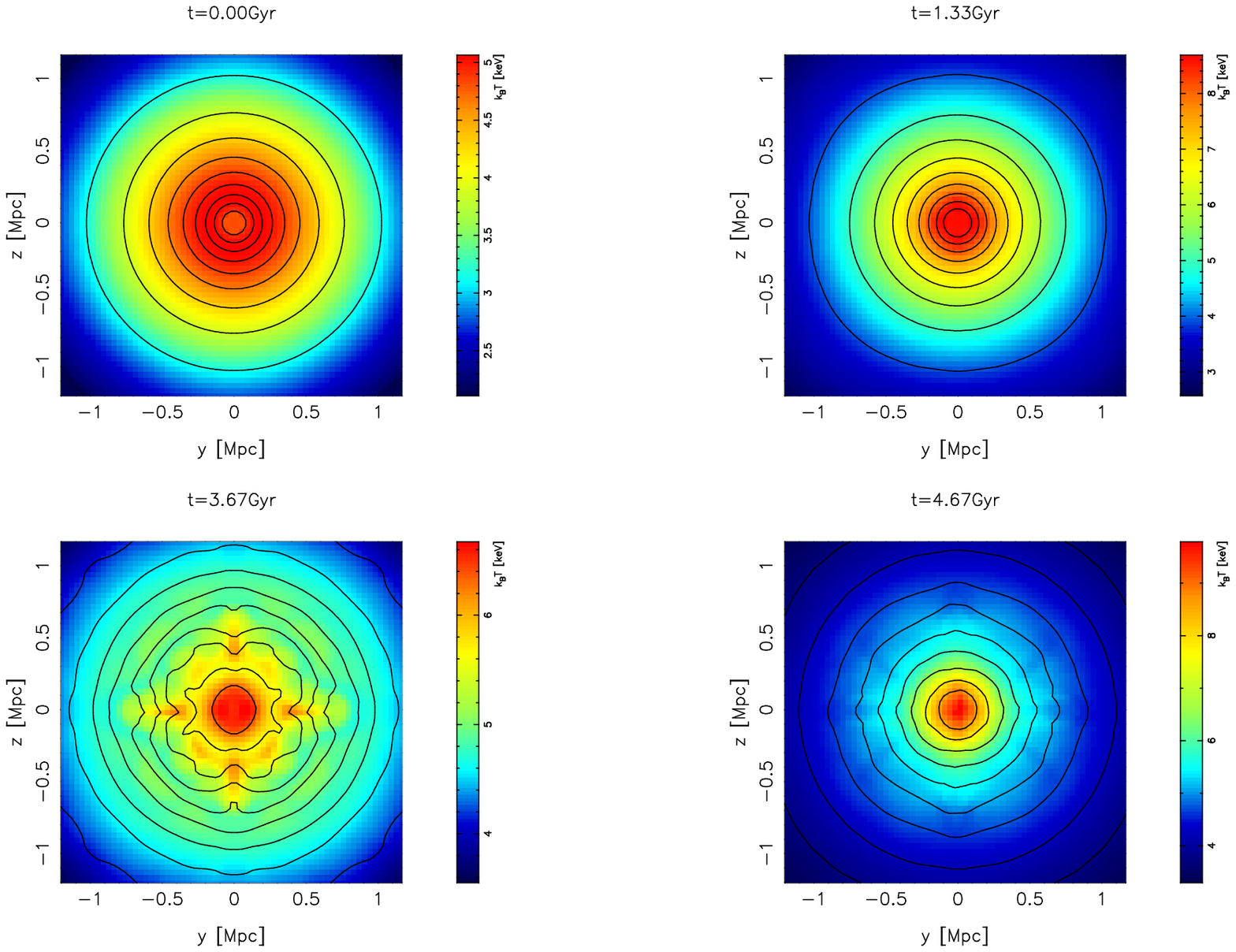}
    %%% \FigureFile(width,height){filename}
  \end{center}
  \caption{Same as figure \ref{fig5}, but seen from the direction parallel
         to the collision axis.}\label{fig6}
\end{figure}

Figure \ref{fig5} and \ref{fig6} present snapshots of the X-ray surface
brightness distribution (contours) overlaid with spectroscopic-like 
temperature one (colors) of the 4:1 merger for the representative epochs
same as in figure \ref{fig3} observed in the direction perpendicular and
parallel to the collision axis, respectively. 
The bow shock is just in the center at
$t=1.33$ Gyr. Kelvin-Helmholtz instability develops and non
axially symmetric structures are clearly seen at $t=3.67$ Gyr when observed
along the collision axis.

\begin{figure}
  \begin{center}
    \FigureFile(160mm,160mm){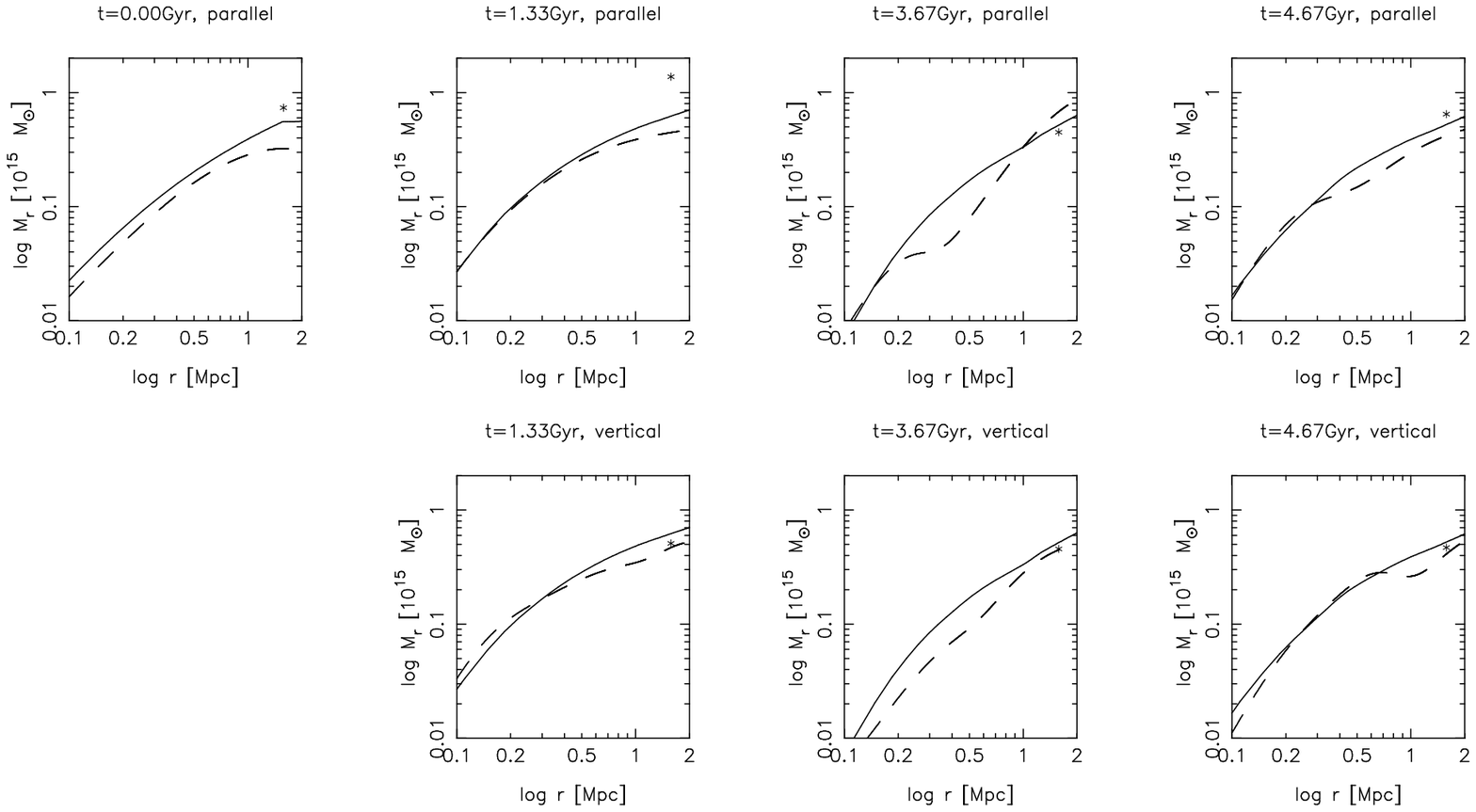}
    %%% \FigureFile(width,height){filename}
  \end{center}
  \caption{Radial profiles of actual mass (solid lines) and estimated mass
         from X-ray data (dashed lines) for the 4:1 merger. 
         Virial mass is also represented
         in asterisks. Upper and lower panels show the mass profiles
         seen from the direction parallel and perpendicular to the collision
         axis, respectively.}\label{fig7}
\end{figure}

\begin{figure}
  \begin{center}
    \FigureFile(160mm,160mm){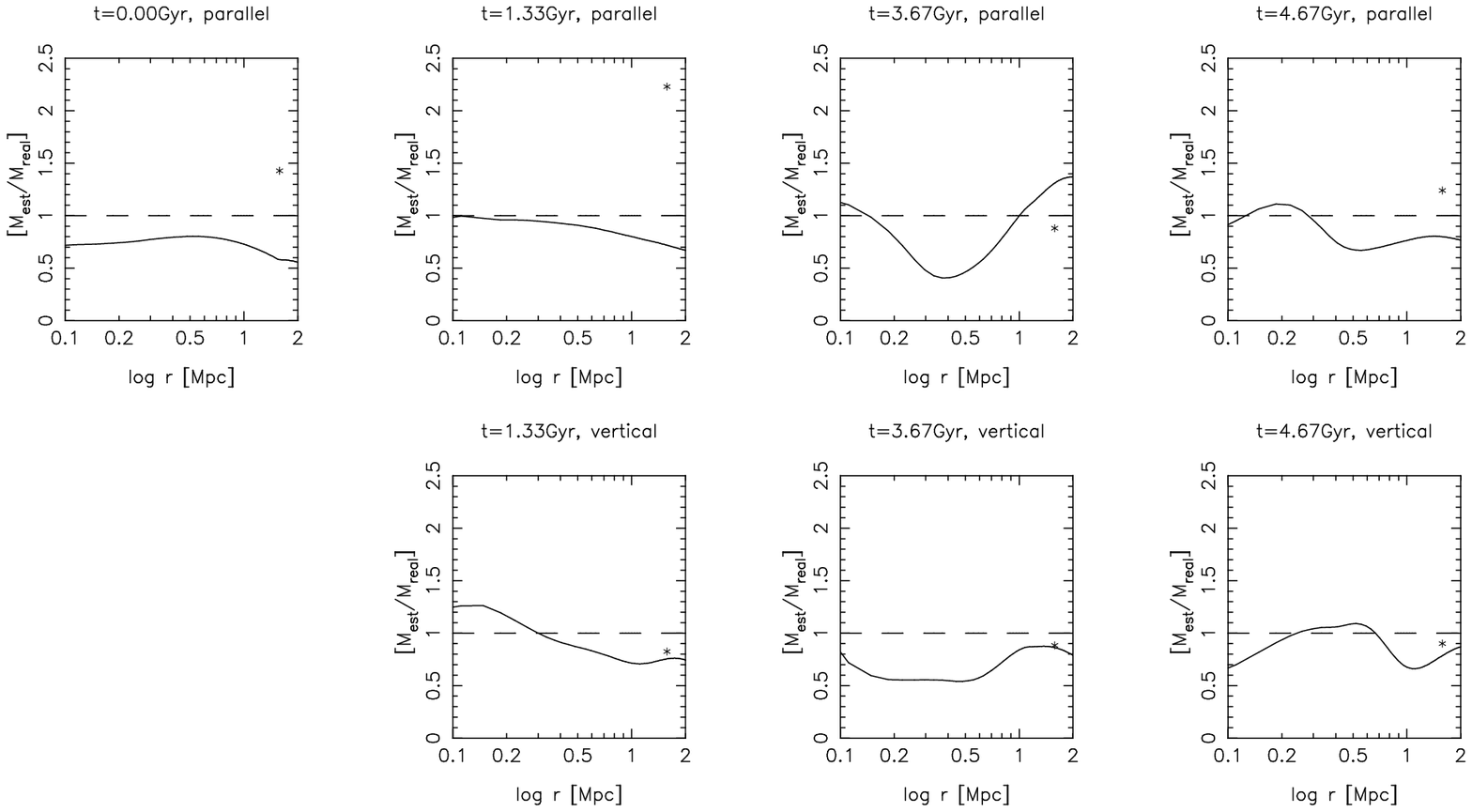}
    %%% \FigureFile(width,height){filename}
  \end{center}
  \caption{Same as figure \ref{fig7}, but for the radial profiles of the ratio
         of estimated mass with X-ray data to the real mass (solid
         lines). The results with Virial mass are also represented
         in asterisks.}\label{fig8}
\end{figure}

First, we compare the estimated mass profiles derived from equation
(\ref{eq:xrmass}) with those of the actual mass.
When calculating an actual mass profile, its center is set at that of the 
X-ray emissivity except for the data at $t=0.0$ Gyr. 
In that case, this definition of the center is not good because the emissivity center is 
located in relatively low density 
region between two subclusters. Therefore, we simply set the mass profile center
at that of the larger subcluster.
Figure \ref{fig7} shows the radial profiles of the actual mass (solid lines)
and estimated mass  (dashed lines) for the data presented in figure 
\ref{fig5} (lower panels) and \ref{fig6} (upper panels). Virial mass is
also presented in asterisks, although it should be noted that the ways to
determine the centers are not common in both the mass estimation methods.
Figure \ref{fig8} shows the same as figure \ref{fig7}, but for the
radial profiles of the ratio of the estimated mass to the real mass.
Again, the ratio of Virial mass is also represented in asterisks.
When the system is observed along the collision axis at $t=0.00$ Gyr,
mass is systematically underestimated at all radii. This is because 
the projected temperature becomes lower owing to the cooler gas of the smaller subcluster.
This results is inconsistent with the virial mass estimation. 
In case of observation along the
collision axis at $t=1.33$ Gyr, while virial
mass estimation results in vast overestimation, X-ray mass estimation 
results in fairly good agreement with the actual mass although the slight
underestimation is seen in the outer region.
Similar trends are seen when observed in the direction perpendicular to
the collision axis at that time, while the both methods lead consistent
results. At $t=3.67$ Gyr, the ratio of the X-ray mass to the real mass shows 
complex and stronger radial dependence when observed along the collision
axis, while the estimation in the direction perpendicular to the axis
results in simple underestimation. 
Negative gradients in the estimated mass profile, which are not 
physically reasonable, are seen in the panels of ``$t=3.67$Gyr, parallel'' 
and ``$t=4.67$Gyr, vertical'' of figure \ref{fig7}. We checked the X-ray image
fitting results and confirmed that these systems were fitted by the double $\beta$-model
and that these structures occurs around the intersection of each $\beta$-model function.
Thus, we conclude that these strange structures in the mass profiles are because of
artifacts of double $\beta$-model fit.

\begin{figure}
  \begin{center}
    \FigureFile(160mm,160mm){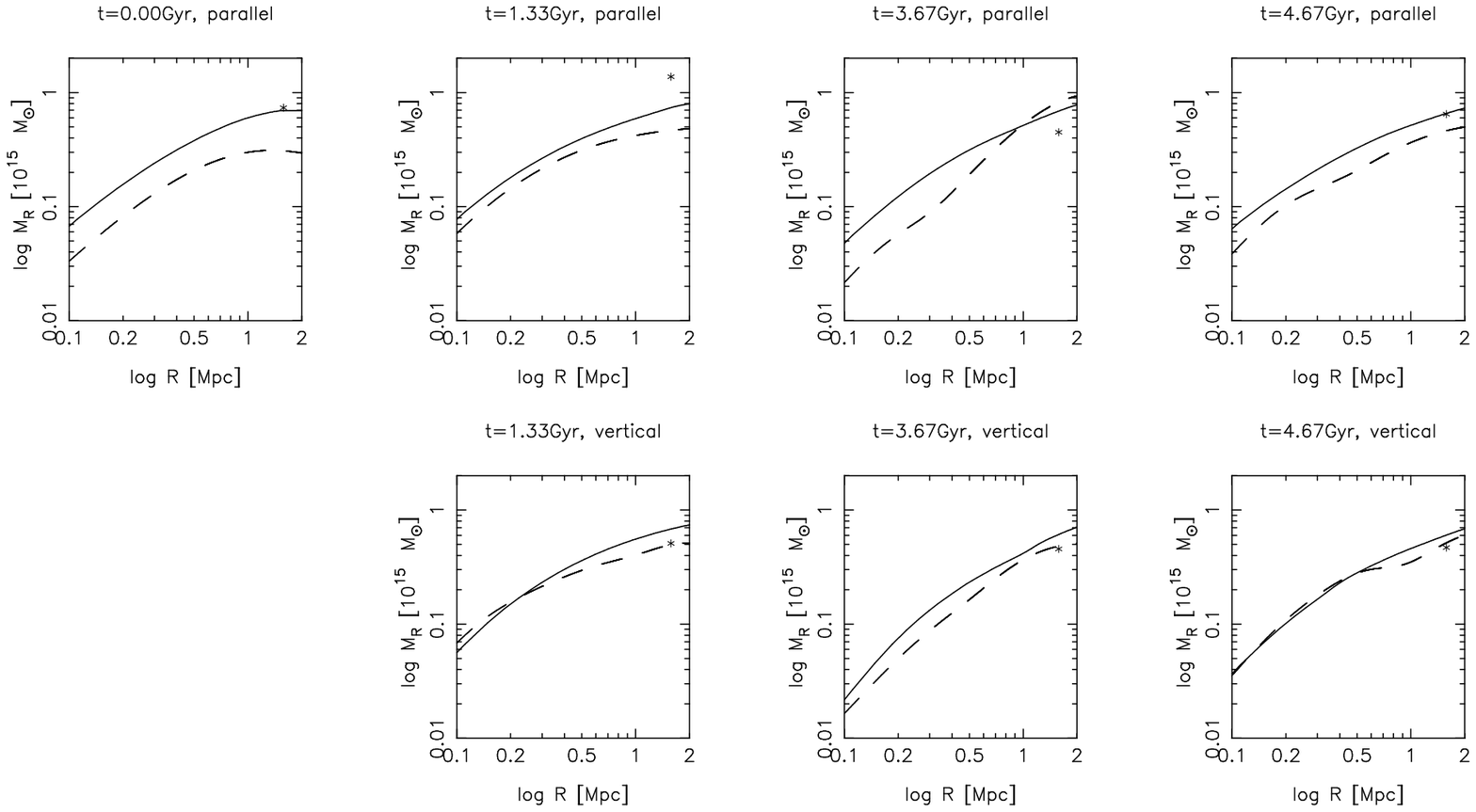}
    %%% \FigureFile(width,height){filename}
  \end{center}
  \caption{Same as figure \ref{fig7}, but for the projected mass profiles
         of actual mass (solid lines) and estimated mass from the X-ray
         data (dashed lines).
         Again, Virial mass is also represented
         in asterisks.}\label{fig9}
\end{figure}

\begin{figure}
  \begin{center}
    \FigureFile(160mm,160mm){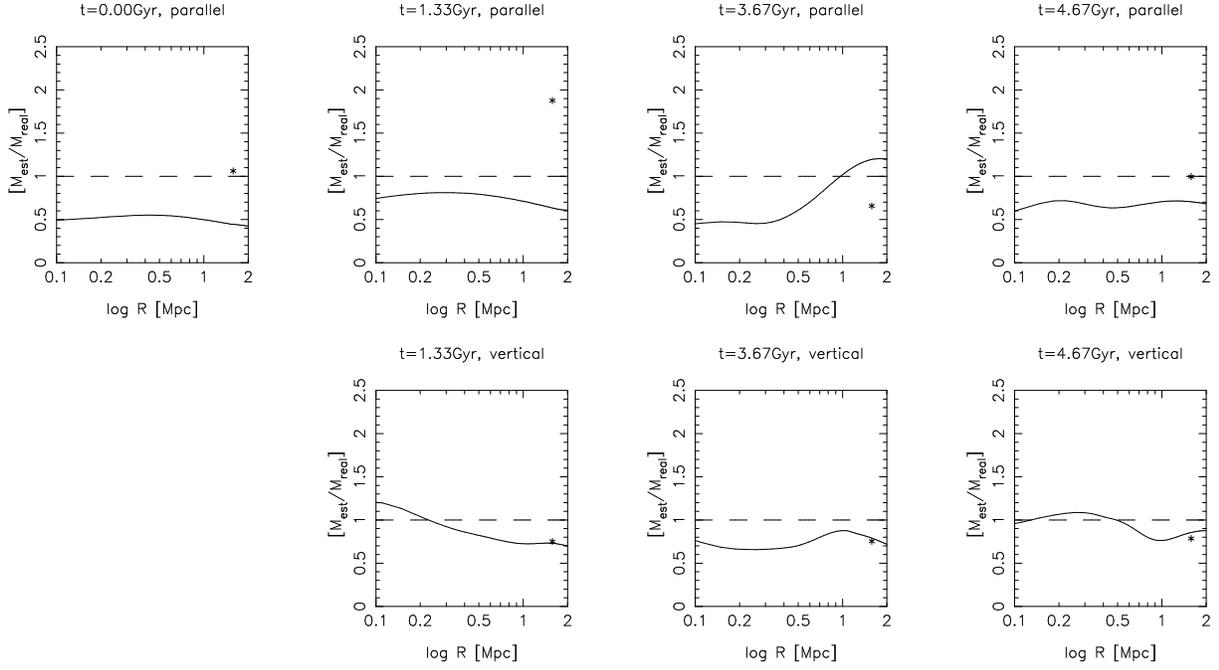}
    %%% \FigureFile(width,height){filename}
  \end{center}
  \caption{Same as figure \ref{fig8}, but for the projected mass.
         The results with virial mass are also represented in asterisks.}\label{fig10}
\end{figure}

It is worthwhile to compare projected mass profiles
calculated from estimated radial mass profiles with real
projected mass profiles, which is directly obtained from gravitational
lensing analysis. The projected mass profile is,
\begin{eqnarray}
  M_{\rm prj}(R) &=& \int_0^R 2 \pi R' \Sigma(R') dR', \\
  \Sigma(R) &=& 2 \int_0^{b_{\rm out}} \rho(\sqrt{R^2+b^2}) db,
\end{eqnarray}
where $\Sigma$ is the surface mass density.
The mass density $\rho(r)$ is calculated from the radial mass profile as follows,
\begin{eqnarray}
  \rho(r) = \frac{1}{4 \pi r^2} \frac{dM}{dr}.
\end{eqnarray}
Figure \ref{fig9} presents the same as figure \ref{fig7}, but for the
projected mass profiles of real mass (solid lines) and estimated mass
(dashed lines). Virial mass is also represented in asterisks. Again, it
should be noted that the centers can be different in both methods for
the same data. Figure \ref{fig10} also shows the same as figure \ref{fig8},
but for the projected mass.
The merging system tends to be elongated towards the collision axis in
three-dimensional mass distribution. As a result, the estimated
project mass under the assumption of spherical symmetry tends to be 
smaller and larger when observed from the direction parallel and
perpendicular to the collision axis, which cause
underestimation and overestimation trends, respectively.
The unphysical structures in mass profiles because of 
the artifacts of double $\beta$ model fit are also slightly recognized in figure \ref{fig9}
though they are much less prominent than in spherical mass profiles.

\subsection{Density and Temperature Profile Models}
\begin{figure}
  \begin{center}
    \FigureFile(160mm,160mm){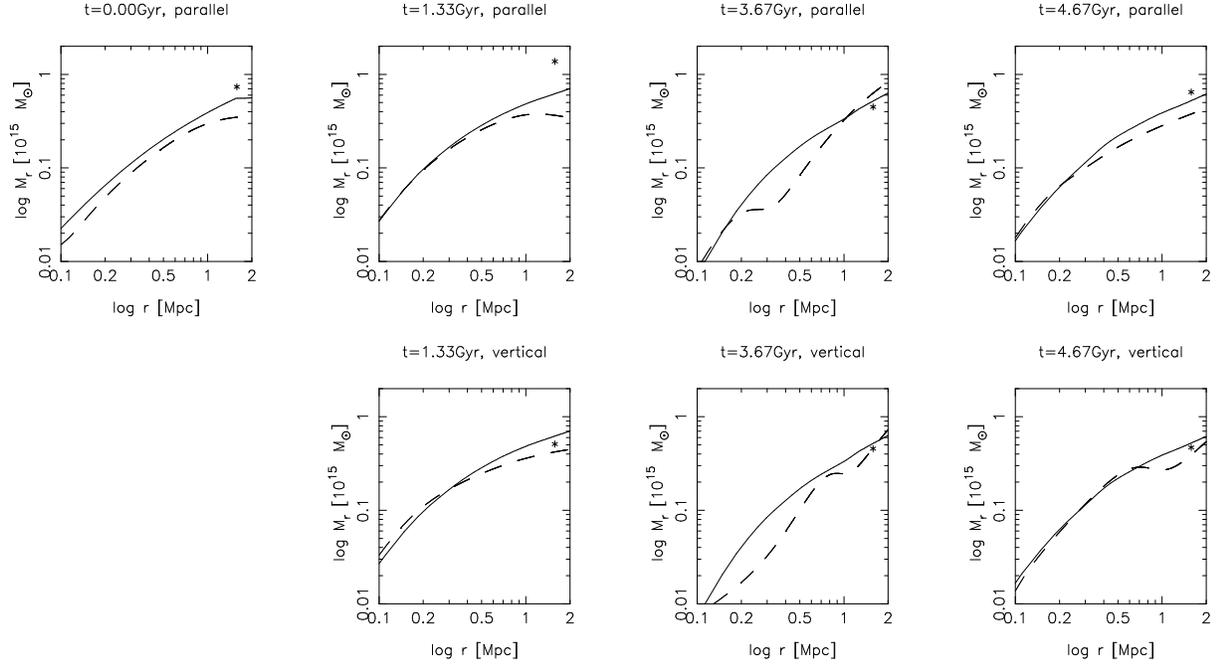}
    %%% \FigureFile(width,height){filename}
  \end{center}
  \caption{Same as figure \ref{fig7}, but for using density and temperature profile models
         proposed by \citet{Vikh06} instead of $\beta$-model.}\label{fig11}
\end{figure}

The mass estimation results can depend on the modeling of density and temperature profiles
\citep{Piff08}. In fact, some of our results show unphysical structures in the mass profile 
caused by the artifact of double beta model fit. 
In order to check this fact we estimate the mass from the X-ray data using 
the density and temperature profile models, which are believed to be more sophisticated,
proposed by \citet{Vikh06} as follows,
\begin{eqnarray}
 n_{\rm e} n_{\rm p} &=& n_0^2 \frac{(r/r_c)^{-\alpha}}{(1+r^2/r_c^2)^{3 \beta - \alpha/2}}
                     \frac{1}{(1+r^{\gamma}/r_s^{\gamma})^{\epsilon/\gamma}}
                    + \frac{n^2_{02}}{(1+r^2/r^2_{c2})^{3\beta_2}},  \label{eq:vikh_den} \\
T(r) &=& T_0 \frac{(r/r_t)^{-a}}{ \biggl [ 1 + (r/r_t)^b \biggr ]^{c/b} }, \label{eq:vikh_tem}
\end{eqnarray}
where we do not use the functional part for the central cool core component in the original form
because clusters used here correspond to so-called non cooling core clusters.
Using these models instead of the $\beta$-model, we estimate a radial mass profile
for the X-ray data of the standard run in a similar way in subsection 3.2. Figure \ref{fig11}
shows the same as figure \ref{fig7}, but for the density and temperature models of equation
(\ref{eq:vikh_den}) and (\ref{eq:vikh_tem}).
In general, the results are quite similar. A negative gradient in the mass profile 
is still seen even with these models.

\section{Summary and Discussion}
\label{s:summ}
We investigate the impact of mergers on the mass estimation of galaxy
clusters using $N$-body + hydrodynamical simulation data. We randomly
select a part of $N$-body particles and recognize them as galaxies
whose line-of-sight velocity is observed. We estimate virial mass from
these data and compare it with actual mass. The results strongly depend
on the observational directions mainly because of anisotropic velocity
distribution. When observed along the collision axis, the mass of the system tends 
to be overestimated. When the smaller cluster mass is larger than a quarter 
of the larger one, the virial mass can be larger than twice of the real mass.
We also make the X-ray surface brightness and spectroscopic-like temperature 
maps from the simulation data. Mass profiles are calculated from these 
``X-ray observational data'' under the assumption of hydrostatic equilibrium.
Again, the estimated mass profiles are compared with those of actual
mass. In general, mass estimation with X-ray data gives us fairly better results
than Virial mass estimation. The dependence upon the observational directions
is weaker, because the gas pressure is isotropic, and because anisotropic 
temperature fluctuations are smoothed out in azimuthal direction.
When the systems are observed along the collision axis, the projected mass tends 
to be underestimated. This fact should be noted when the Virial and/or 
X-ray mass are compared with gravitational lensing results.

\subsection{Mass Estimation Uncertainty and the Morphology of X-ray Image}

It is useful to discuss the relationship between the mass estimation 
uncertainty and the morphology of X-ray image. 
One may easily guess that 
clusters with an extremely elongated X-ray image 
(e.g., the snapshot at t=3.67 Gyr in figure \ref{fig5})
are not in hydrostatic equilibrium. Indeed, the mass is systematically 
underestimated in this system (see ``t=3.67Gyr, vertical'' panel in Fig8).
Roughly speaking, the X-ray mass estimation results in underestimation of 20-40 \%.
In other word, mass can be estimated as accurately as within a factor of two 
even for such an irregular cluster. The estimation accuracy becomes better
for the projected mass comparison (e.g., the snapshot at t=3.67 Gyr 
in figure \ref{fig10}). This is because the underestimation trend caused by 
irregular morphology and bulk flow motion is partly compensated
by the overestimation trend caused by the projection effect.

\begin{figure}
  \begin{center}
    \FigureFile(160mm,160mm){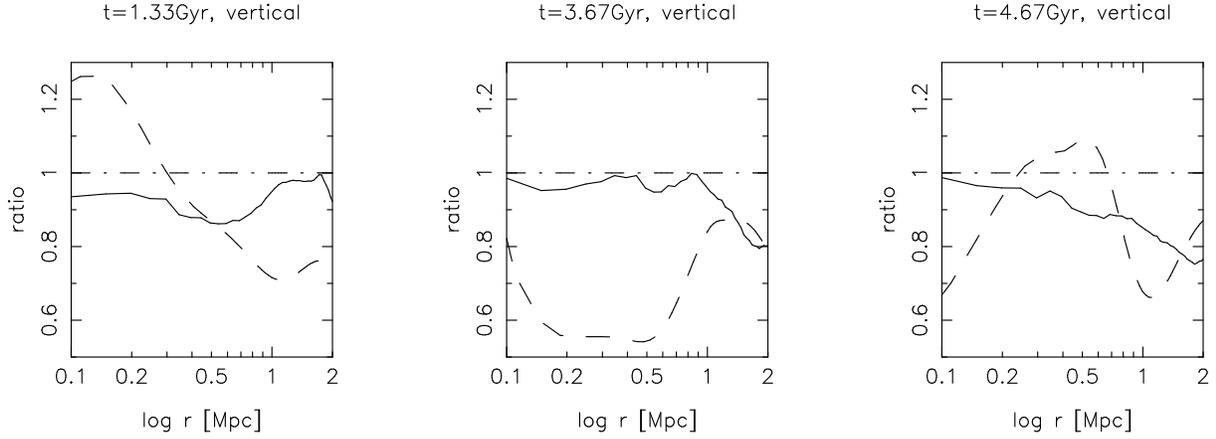}
    %%% \FigureFile(width,height){filename}
  \end{center}
  \caption{The axial ratio of the X-ray surface brightness as a function of the aperture radius (solid line)
         and ratio of estimated mass with X-ray data to the real mass (dashed lines) for the standard run
         seen from the direction perpendicular to the collision axis.}\label{fig12}
\end{figure}

To quantify the image morphology, we use the axial ratio of the surface brightness
image defined through the second moments of the surface brightness,
\begin{eqnarray}
   M_{ij} = \sum I_X x_i x_j,
\end{eqnarray}
where the summation is conducted over the coordinates whose origin is at the X-ray peak
and within a certain aperture centered on the origin \citep{Ohar06}. 
Our models have axial symmetry for the collision axis ($x$-axis).
Therefore, we only need to calculate $M_{11}$ and $M_{22}$ and do not have to diagonalize
the matrix $M$ for the X-ray images seen from the direction perpendicular to the collision axis, 
when we choose the $x_1$ and $x_2$ axes to be parallel to the $x$ and $y$, respectively. 
We actually calculate $M_{12}(=M_{21})$ and confirm that $|M_{12}/M_{ii}|$ is typically $\sim  0.005$.
Then, the axial ratio $\eta$ is defined as,
\begin{eqnarray}
   \eta = \Bigg\{
          \begin{array}{cc}
              M_{11}/M_{22}, & M_{11} \le  M_{22}, \\
              M_{22}/M_{11}, & M_{11} > M_{22}.
          \end{array}
\end{eqnarray}
Thus, the axial ratio lies between 0 and 1 by definition, and $\eta=1$ means a circular cluster. 
In some clusters with complex morphology, such as ones with three or more peaks, 
the axial ratio is not a good indicator \citep{Vent08}. However, because our model clusters tends to have 
a relatively simple elliptical shape, it does not matter.
Figure \ref{fig12} shows the $\eta$ as a function of the aperture radius (solid lines) and the ratio
of estimated and real mass (dashed lines) for the standard run seen from the direction perpendicular 
to the collision axis.
It is interesting that the inner region is nearly circular ($\eta \simeq 1$) but mass estimation error is 
relatively large at $t=3.67$ and $4.67$ Gyr. This means that the morphology determined from only inner region
is not a good indicator of mass estimation error, which should be cared especially 
for the distant faint clusters.

On the other hand, it should be noted that clusters with spherical and 
regular X-ray morphology do not always result in good mass estimation.
This is especially true for head-on merger clusters observed nearly 
along the collision axis. Spherical mass can be both overestimated and 
underestimated, depending on the radial distance and the phase of merger,
although the results tend to be more accurate than in case of virial mass
(see upper panels in figure \ref{fig8}). About $\sim 20$ \% errors
are typically seen. Even nearly 50 \% errors can be found 
in the ``t=3.67Gyr, parallel'' panel in figure \ref{fig8}.
As for the projected mass results, however, the situation is a little 
bit different, where the X-ray mass is systematically smaller 
than the real mass (see the upper panels in figure \ref{fig10}).
In these cases, $\sim$ 50 \% underestimations can occur.
In both cases, anyway, it should be noted that 
X-ray and virial mass estimations give us inconsistent results, 
and that both are significantly different than the real mass,
although the X-ray image is quite symmetric.
This implies that these clusters do not follow the correlation
between X-ray temperature and galaxy velocity dispersion 
for relaxed clusters. In other words, these clusters show a peculiar 
$\beta_{\rm spec}$ value, which has been already recognized as a good indicator
for merging clusters \citep{Ishi96, Taki00}. 
These results can qualitatively explain the discrepancy between 
X-ray and lensing mass reported in CL 0024+17 \citep{Ota04, Jee07}.

Recently, \citet{ZuHo09} studied a line-of-sight galaxy cluster collision
using numerical simulations, and compared their results with Cl 0024+17
in detail. Their results are basically consistent with ours, but
there are several differences. As seen in figure \ref{fig10}, 
systematic underestimation trends are clearly seen in the projected mass 
when observed along the collision axis, which is qualitatively consistent with theirs. 
As written in subsection 2.2 of \citet{ZuHo09}, their collision velocity is approximately 
twice of the free fall velocity. This is an extremely rare case in the CDM universe though it might be good 
for reproducing the Cl 0024+17 results. On the other hand, the collision velocities 
in our simulations are comparable to but a little bit smaller than the free fall one, 
which means a more usual case. Thus, it is not surprising that \citet{ZuHo09} shows more 
serious underestimation, and that their results are quantitatively different than ours.
In addition, though both \citet{ZuHo09} and our models used the double-beta model fitting, 
its interpretation and resultant procedure of mass estimation are different. 
\citet{ZuHo09} assumes that double-beta surface brightness distribution means two individual 
spherical clusters aligned along the line-of-sight. Therefore, they calculate hydrostatic 
mass for two spherical clusters and sum them in deriving projected mass. 
On the other hand, we assume that the double-beta surface brightness distribution means 
a single spherical cluster with a relatively complex density profile that cannot be well represented 
by a single beta-model \citep{Ikeb04}. Thus, we calculate hydrostatic mass for a single spherical cluster.

\subsection{ICM Internal Velocity}
It is quite natural that ICM internal velocity as well as member galaxies'
velocities can be another probe of merger dynamics and mass estimation uncertainty.
In addition, although subclusters are in hydrostatic equilibrium 
in the initial conditions in our merger simulations, it is most likely that clusters 
before major mergers have already undergone some mergers in the CDM universe. 
As a result, the merger progenitors could have significant bulk and/or turbulent 
flow motions in the ICM. This causes additional underestimation 
in mass measurement with X-ray data \citep{Fang09}.
Anyway, firm direct detection of ICM motion is highly desired in this regard
though this is not an easy task in the current status of X-ray astronomy
(e.g. \cite{Ota07,Fuji08,Suga09}).
The ASTRO-H satellite (former NeXT; e.g. \cite{Taka06}),
which is planned to be launched around 2013, will enable us 
to measure the line-of-sight velocity directly with the X-ray microcalorimeters and 
provide us with useful information on the dynamical status of the ICM. 
Another possible observational approach to the ICM
motion is the Sunyaev-Zel'dovich (SZ) effect. In principle, the line-of-sight and 
tangential velocity components can be measured with the kinematic SZ effect and 
SZ polarization, respectively \citep{Suny80}, 
though they are still challenging in the present status of the observations.

\subsection{Member Galaxy Distribution}
As mentioned above, we sampled particles corresponding to 
the member galaxies from $N$-body particles in virial mass estimation.
This means that the phase space distribution of the galaxies
are assumed to be essentially the same as that of DM. 
However, it is likely that this assumption is not so good in realistic situations.
Using cosmological simulations that include cooling and star formation, \citet{Naga05} showed that 
the concentration of the radial distribution of simulated galaxies is lower than that of dark matter
even for relaxed systems, thereby implying errors in the mass estimation through the virial theorem. 
On the other hand, our aim is to evaluate errors caused by dynamical motions during mergers, therefore,
for systems that are far from dynamical equilibrium. We conclude that our findings about
virial mass estimates are not significantly affected by such biases \citep{Naga05}.


%% Included in this acknowledgments section are examples of the
%% AASTeX hypertext markup commands. Use \url without the optional [HREF]
%% argument when you want to print the url directly in the text. Otherwise,
%% use either \url or \anchor, with the HREF as the first argument and the
%% text to be printed in the second.

\bigskip
M. T. would like to thank N. Okabe, M. Takada, and M. Nagashima
for helpful comments. The authors are grateful to an anonymous referee
for his/her useful comments and suggestions which improved the manuscript.
Numerical computations were carried out
on VPP5000 and XT4 at the Center for Computational Astrophysics, CfCA, of the
National Astronomical Observatory of Japan.
M. T. was supported in part by a Grant-in-Aid from the
Ministry of Education, Science, Sports, and Culture of Japan (16740105,
19740096).

%%%%%%%%%%%%%%%%%%%%%%%%%%%%%%%%%%%%%%%

\appendix
\section*{Evolution of the Equilibrium Model Cluster}
\begin{figure}
  \begin{center}
    \FigureFile(160mm,160mm){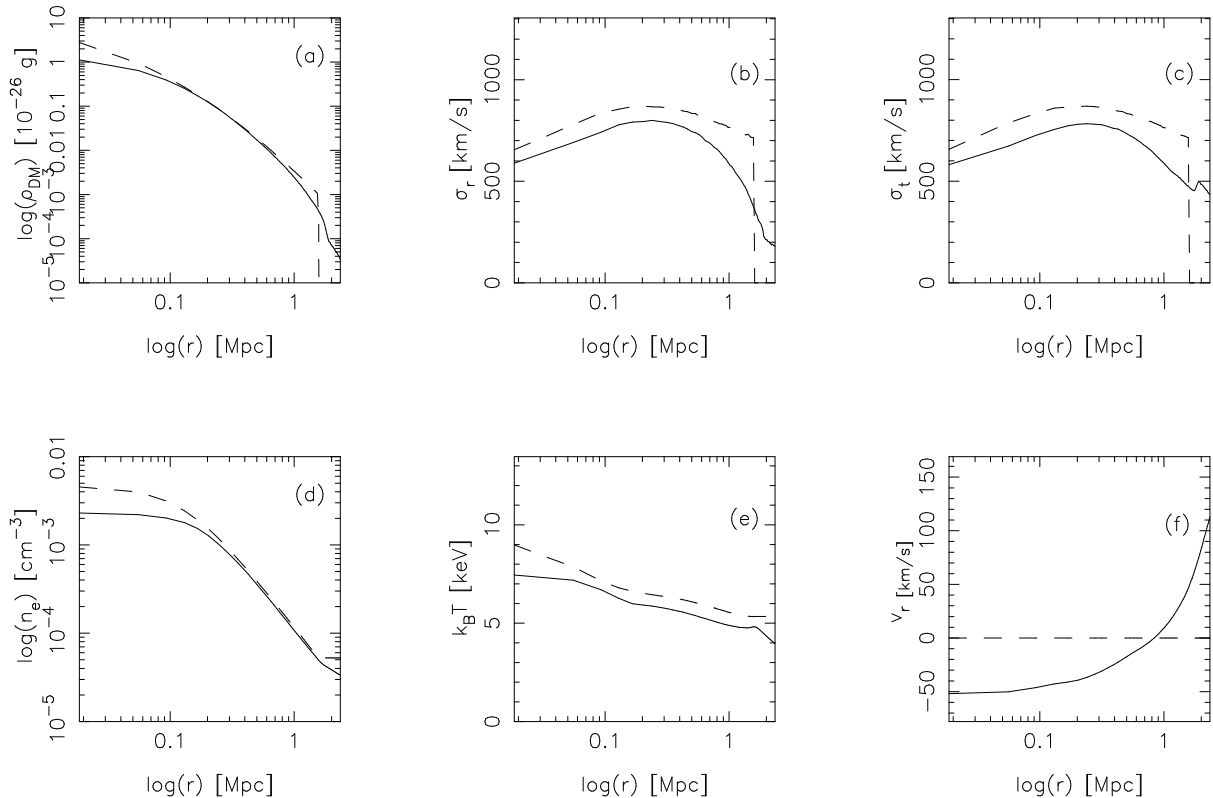}
    %%% \FigureFile(width,height){filename}
  \end{center}
  \caption{The radial profiles of (a) DM density, (b) standard deviation of DM radial velocity,
         (c) that of tangential velocity, (d) electron number density of ICM, (e) ICM temperature, and
          (f) radial velocity of ICM for the equilibrium cluster model. The dashed and solid lines show
         the profiles at $t=0$ and $4$ Gyr, respectively.}\label{fig13}
\end{figure}

Our initial cluster model is in dynamical equilibrium when the spatial distribution of DM and
ICM extend to the infinity. Strictly speaking, this is not the case for the actual simulations.
In addition, finite spatial resolution of the code could affect the structure and evolution especially
in the central region. To check these issues, we perform the run of a single cluster where the initial
conditions are the same as those of the larger cluster in the standard run.
Figure \ref{fig13} shows the radial profiles of various physical values, where the dashed and solid 
lines represents the profiles at $t=0$ and $4$ Gyr, respectively. The DM density profile in the central 
region becomes slightly flatter but still has the cuspy structure. In the outer region, 
DM density profile change is much less significant. The shape of the DM velocity dispersion profiles 
does not change significantly for both radial and tangential components and the velocity distribution
keeps to be isotropic, although the absolute value of the velocity dispersions become slightly smaller.
Similar trends are seen in the ICM radial density and temperature profiles. The ICM has finite radial 
velocity at $t=4$ Gyr, which is much less than the sound velocity. 

It is certain that our initial model shows some signs of evolution quantitatively. 
However, its impact for mass estimation is very limited. For example, typical values of the excited radial 
velocity of the ICM ($\sim$ 50 km s$^{-1}$) is much smaller than the sound velocity of 7 keV ICM 
($\sim$ 1400 km s$^{-1}$), as in shown figure \ref{fig13}. 
This means that the kinetic energy of the ICM is only 
$\sim 0.1 \%$ of the thermal one. As for the dark mater component, we actually checked the results of 
Virial mass estimation for the data shown in figure \ref{fig13}, and confirmed that the errors becomes 
only less than $\sim 10\%$, which is smaller than typical statistical errors with $N_{\rm samp}=100$. 
These effects are clearly smaller than the impact of merger itself. Therefore, we think that our initial 
model is suitable enough for the main purpose of this paper.

We actually followed the time evolution of the isolated initial model cluster
for 5 Gyr. However, we did not see transition to another equilibrium configuration. 
Even if we possibly find another equilibrium configuration after 
the much longer run, we do not think that the configuration is much
preferable for the initial conditions, considering that it could be significantly
altered from the NFW profile that represents the structure of dark halos found in
cosmological simulations reasonably well.

%%%
% See the manual for the detail.
%%%
%\begin{thebibliography}{}
% Journals(e.g. A\&A,ApJ,AJ,NMRAS,PASP ...)
% Authors, Year, Journal, Vol#, Page#
% Journal Title Abbreviation >> http://www.asj.or.jp/pasj/Jabb.html
%\bibitem[Aauthor et al.(2001)]{key-1}
%   Aauthor, A., Bauthor, B., Cauthor, C.\ 2001, PASJ, vol, page
% Books
%\bibitem[Aauthor \& Author(2001a)]{key-2}
%   Aauthor, A., Author, B.\ 2001, Name of Book(Publisher, Tokyo) ch0
% Books
%\bibitem[Aauthor \& Bauthor(2001b)]{key-3}
%  Aauthor, A., Bauthor, B.\ 2001, Name of Book(Publisher, Tokyo) page0
%......
% Editorial Books
%\bibitem[Dauthor(2001)]{key-n}
%  Dauthor A. A.\ 2001, in Name of Book,
%   ed Editor D.\ Editor(Publisher, Tokyo) page0
%\end{thebibliography}
%%%

\end{document}